PAPER

# Graphene-like nanoribbons connected by four-/five- membered rings on pentacene/picene precursors, Au(110) surface

Yu Yao [a], Yu Wang [a], Shuangxiang Wu[a], Qiang Zhang [a], Yifan Wang [a c], Hui Zhang * [a b c]

The rapid development of fabrication of functional high-quality graphene-like nanoribbons has increasingly reliant on integration of diverse nanofabrication platforms. The demand for innovative techniques to enable the exploration and precise manipulation of graphene-like nanoribbons. Herein, we present an on-surface synthesis approach for graphene-like nanoribbons, achieved through the self-assembly and annealing of pentacene/picene monomers on an one-dimensional(1D) Au(110) substrate. Scanning tunneling microscope(STM) analysis reveals that the formation of four- and eight- membered rings between adjacent molecules. Furthermore, we demonstrate a technique for manipulating pentacene dimer without breaking structural integrity by STM tip manipulation. Our results exhibit a potential platform for the development of next-generation graphene-based quantum computing architectures, as well as a technique for fabricating functional graphene-like nanoribbons with exceptional precision.

## Introduction

Among substrate-supported materials, graphene nanoribbons have emerged as a prolific platform for investigating novel two-dimensional materials due to the unique nanoscale physical properties[1,2], such as spin-polarized edge states[3], giant magnetoresistance[4], the edge quantum confinement effect[5,6], and strong electron correlation[2]. Benefited from a wide range of artificially tuneable properties, graphene nanoribbons have demonstrated their applicability in the fabrication of cutting-edge nanodevices, such as field-effect transistors[7], cascaded spintronic logic circuits[8], width-modulated heterojunctions[26] and strong photoluminescence emitters[27]. These advancements underscore their exceptional potential as building blocks for next-generation nanotechnology. Currently, the atomically precise synthesis of custom-designed graphene-like nanoribbons[9,10,11] has garnered significant interest. From a technical perspective, this approach increasingly demands the development of meticulously designed graphene-like nanoribbons to meet the growing requirements.

Traditional approaches to achieving this include chemical wet etching[12], electron beam etching[13], and unzipping of carbon nanotubes[14]. However, traditional methods encounter significant challenges in enhancing the diversity and precise manipulation of graphene-like nanoribbons at the atomic scale, primarily due to limitations in nanofabrication techniques. Currently, commonly employed methods include tip-induced selective bond chemistry[23], on-surface polymerization[24], and molecular self-assembly[25]. These approaches mainly depend on the precise positioning of the STM tip or the application of external field excitation such as thermal or electric fields. While the controllability of these methods is limited, and they are highly susceptible to substrate interactions and temperature fluctuations, posing significant challenges for their implementation in future industrial applications.

Over the past decade, substrate-assisted catalytic synthesis methods[28,29,30,31,32,33,34] have garnered significant attention due to their enhanced stability, greater configurational control and higher precision. These methods exploit functionalized precursor molecules (e.g., halogenated functional groups) to enable selective dehydrogenative cyclization at lower temperatures, facilitating the formation of molecular chains with exceptional configurational precision. However, the applicability of molecular precursors and substrates remains limited, particularly when employing templates such as the Au(110) surface, with the underlying mechanisms of these constraints still unclear. Consequently, it is imperative to develop substrate-assisted, covalent-linkage-based synthesis strategies to achieve improved stability, broader configurational tunability, and deepen the understanding of graphene-substrate interaction.

Here, we report a strategy to modulate the structure of graphene-like nanoribbons during their preparation on a catalytic substrate via inducing covalent interlinking of the precursor molecules. Following this approach, we successfully fabricated multiple graphene-like junctions with one-benzene-ring width on a catalytic substrate. Through the cyclo-dehydrogenation of precursors, pentacene molecules

[a] *International Center for Quantum Design of Functional Materials (ICQD), Hefei National Research Center for Physical Sciences at the Microscale, University of Science and Technology of China, Hefei 230026, China.*
[b] *Hefei National Laboratory, University of Science and Technology of China, Hefei 230088, China*
[c] *CAS Key Laboratory of Strongly-Coupled Quantum Matter Physics, and Department of Physics, University of Science and Technology of China, Hefei 230026, China*

*Email: huiz@ustc.edu.cn*

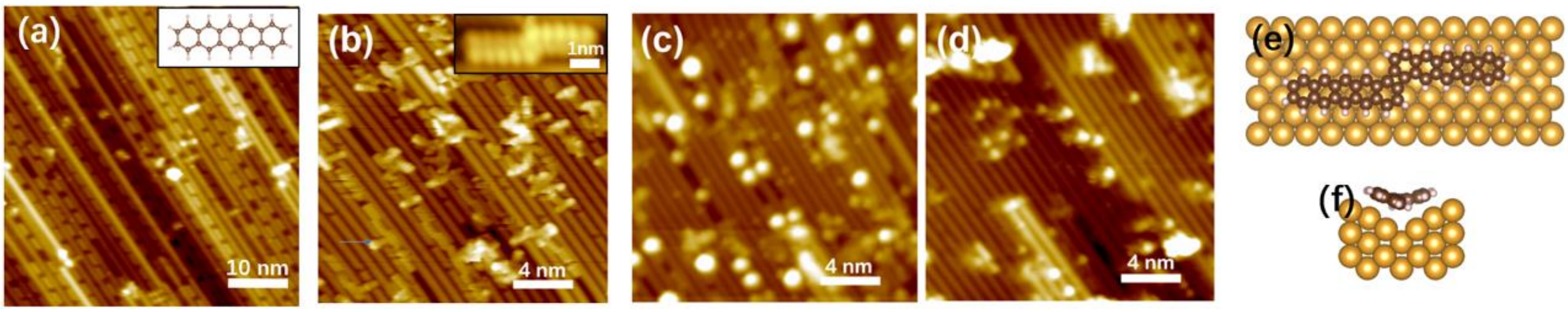

**Fig. 1** On-surface synthesis of four-membered ring formation on the Au(110) substrate. (a) STM topography of 0.8 ML pentacene molecules deposited on the Au(110) substrate at room temperature ($V_s$ = −1 V and $I$ = 1 nA). The inset shows the structure of pentacene monomer. (b) 0.8 ML pentacene self-assembly after annealing at 638 K for 1 hour. The inset shows the detail STM topography image of the pentacene dimer. (c) After annealing at 638 K for 2 hours. (d) After annealing at 673 K for 1 h. (e) Front view of a pentacene dimer connected by a four-membered ring. (f) Side view of a pentacene dimer connected by a four-membered ring.

were polymerized into 1D graphene-like nanochains featuring armchair-shaped edges. In contrast, we found that picene molecules failed to achieve precise confinement within the grooves of the Au(110) substrate during the annealing process, leading to the formation of multiple molecular orientations. Furthermore, we successfully manipulate pentacene and picene chains by folding them with an STM tip without causing structural damage, demonstrating that these nanostructures maintain their mechanical integrity during manipulation. This method provides a pathway for fabricating GNRs with diverse configurations, paving the way for the development of future quantum computing devices based on graphene molecular systems.

## Material preparation

Sample preparation and characterization were performed in an ultrahigh-vacuum (UHV) system equipped with a commercial low-temperature scanning tunneling microscope (STM) (Createc, Germany) operating at a base pressure below $2 \times 10^{-10}$ mbar. The Au(110) surface was prepared through repeated cycles of $Ar^+$ sputtering (1 kV, 20 min) followed by thermal annealing at 673 K for 30 min. Prior to deposition, pentacene (Sigma–Aldrich, purity > 99.9%) and picene (TCI, purity > 99.8%) molecules were degassed for over 10 hours. The molecules were subsequently deposited onto the substrate at room temperature with one monolayer (ML) thickness. All STM images were recorded at 78 K using a tungsten tip for both imaging and manipulation of the nanoribbons.

## Self-assembly of the pentacene dimer with a four-membered ring

By precisely controlling the deposition rate of pentacene molecules on the re-annealed Au(110) $1 \times 2$ substrate at 473 K, we successfully obtained 0.8 ML self-assembly of pentacene on the Au(110) surface (Fig. 1a). Most pentacene molecules adsorbed onto the surface in a lying-down configuration, self-assembled into well-ordered quasi-1D chains, consistent with previous reports[16]. The molecular arrangement induced a structural phase transition in Au(110) surface, changing from $1 \times 2$ reconstruction to $1 \times 3$ reconstruction[17,18]. To investigate the self-assembly mechanism of pentacene, we employed substrate temperature and annealing time as independent variables and analyzed their influence on the mechanical behavior of pentacene. Upon annealing at 623 K for 2 hours, some pentacene molecules diffused along the grooves of the Au(110) surface due to the weak interaction between the molecules and the substrates (Fig. 1b). Approximately 30% of the molecules underwent chemical bonding, forming linear chains. With prolonged annealing time, pentacene molecules exhibited gradual desorption from the substrate, accompanied by a confined migration within the substrate grooves. This behavior indicates a competitive dynamic process between molecular diffusion and covalent linking. The observed molecular chain exhibited lengths corresponding to 2 to 5 pentacene monomers. Continue to increase the substrate temperature (Fig. 1c) caused pentacene desorption to occur at 638 K. Subsequently, extending the annealing time independently led to approximately 40% of pentacene molecules undergoing random diffusion and forming clusters (Fig. 1c). Finally, when the annealing temperature reached 673 K (Fig. 1d), over 80% of pentacene molecules chemically desorbed from the Au(110) surface, and the Au(110) substrate reverted to its $1 \times 2$ reconstruction. In summary, higher substrate temperatures resulted in pentacene desorption, whereas prolonged annealing time promoted random molecular diffusion and the formation of pentacene clusters.

## Multi-configurations of the pentacene dimer

From the insert image of Fig. 1b, a well-defined one-dimensional pentacene dimer emerged at an annealing temperature of 633 K. With prolonged annealing time, it became apparent that two neighbouring pentacene molecules underwent polymerization to form the pentacene dimer. In Fig. 1b, a contrast between the inner and outer edges of each pentacene monomer is observed, indicating that the pentacene dimer does not lie perfectly flat within the Au trough. Instead,

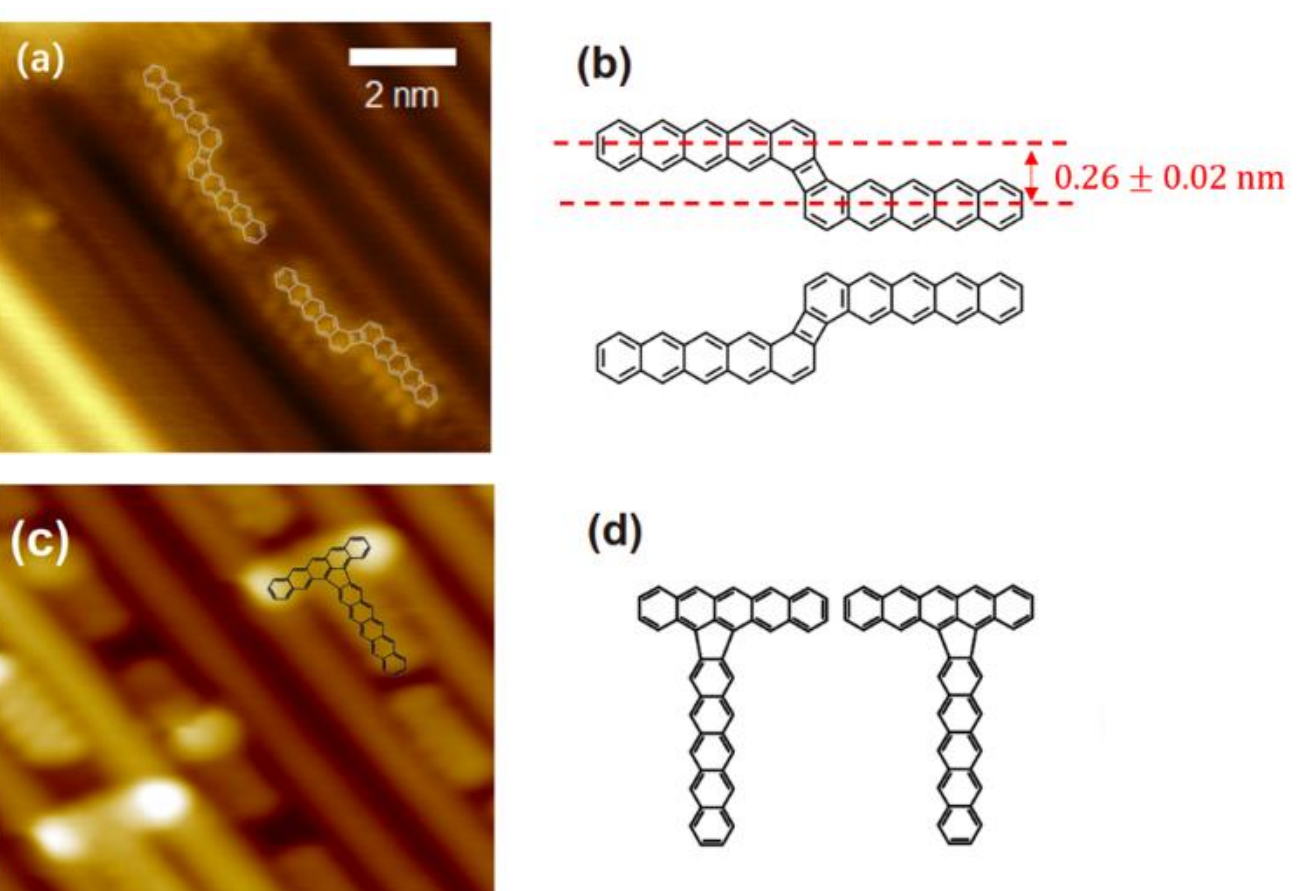

**Fig. 2** (a) STM topography and (b) structural representation of syn-conformal (upper) and anti-conformal (lower) pentacene dimer structures. (c) STM topography and (d) structural representation of syn-conformal (left) and anti-conformal (right) T-shaped pentacene dimers ($V_s$ = −1.2 V and $I$ = 1.1 nA) each pentacene molecular segment aligns with the opposing walls of the Au(110) reconstruction grooves, resulting in the entire pentacene dimer adopting a V-shaped configuration (Fig.1e and f).

We focused on studying the carbon four-membered ring positioned between two Au channels. By measuring the distance between the central chain axes of two neighbouring pentacene monomers, we obtained a value of 0.26$\pm$0.02 nm (Fig. 2b), which aligns closely with results from density functional theory calculations[21]. Furthermore, the measured total length of the dimer matches precisely with the combined lengths of two individual pentacene molecules and one carbon four-membered ring. STM topography images further revealed the characteristics of the carbon four-membered ring, showing its interconnection with pentacene segments through $sp^2$ hydrogenation. This $sp^2$ hydrogenation fosters the transformation of the entire chain into a conjugated $\pi$-bonded hydrocarbon, distinctly different from carbon chains with C-H single bonds, which typically produce protrusions at monomer connections[16]. From the STM image (Fig. 1b), no such protrusions are observed in either the pentacene monomers or the carbon four-membered ring, suggesting the potential coexistence of the carbon four-membered ring within the same plane shared by the two pentacene monomers.

Theoretically, we attribute the formation of the carbon four-membered ring to varying dehydrocyclization reactive sites on the trailing benzene ring. In pentacene dimers, the carbon four-membered rings primarily exhibit two configurations: syn-conformational (Fig. 2b, upper molecule) and anti-conformational (Fig. 2b, lower molecule). Both configurations consist of a pair of chiral molecules with symmetrical structures. Furthermore, the induced dehydrogenation between chiral pentacene dimers facilitates the forming of a 1D graphene-like nanochain with a single-benzene-ring-width structure.

During the formation of pentacene dimer, the Au(110) substrate played a pivotal role in two aspects. First, the gold substrate served as a template, immobilizing the molecules and catalyzing the dehydrogenation process between adjacent pentacene monomers. Second, the hydrogen atoms at the end of pentacene molecules facilitated mutual bonding, promoting C-C $sp^2$ hybridization[16]. Notably, previous studies reported that pentacene molecules are unlikely to form connections via C-H single-bond involving $sp^3$ hybridization[19,20]. In our experiments, over 80% of the molecules connected through carbon four-membered rings formed dimers, trimers and other chain-like polymers (Fig.2).

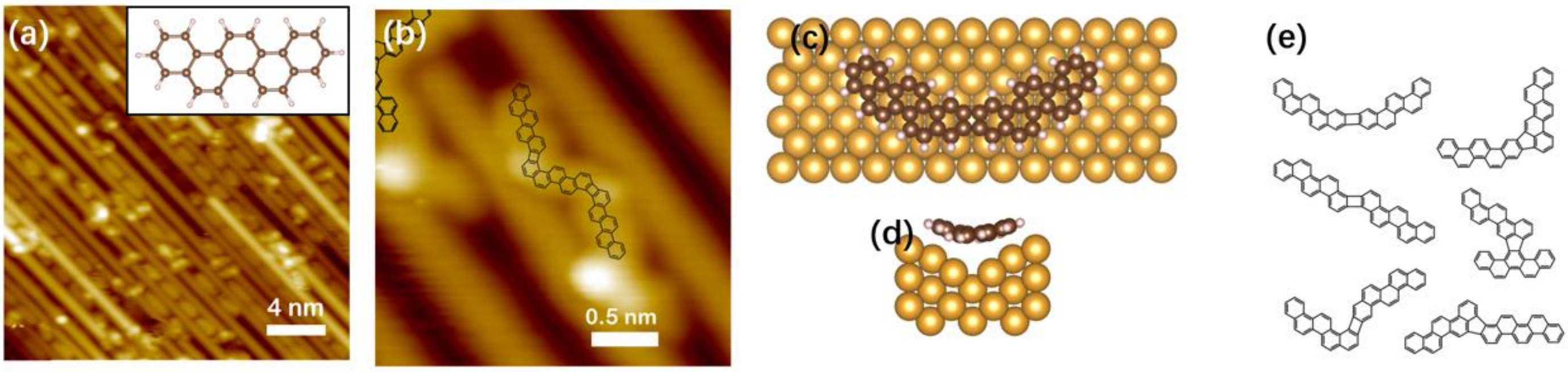

**Fig. 3** (a) STM topography showing dimer configurations with different connection ($V_s$ = −0.8 V and $I$=1 nA), the inset shows the structure of a picene monomer. (b) STM image of picene dimer on Au(110) substrate after annealing at 573 K for 2 hours. (c) Front view of a picene dimer connected by a four-membered ring. (d) Side view of a picene dimer connected by a four-membered ring. (e) Observed configurations of picene dimers.

In addition to the linear pentacene dimer, we identified a distinct type of T-shaped pentacene junctions. To understand the formation mechanism of these T-shaped junctions, we analysed the evolution of pentacene molecules during the synthesis. STM data revealed the following sequence of events: initially, some pentacene molecules were deflected by the substrate and oriented perpendicular to the direction of the Au chain. As the annealing process advanced, these perpendicular pentacene molecules combined and aligned with the molecules transversely within the Au chains. Ultimately, through controlled dehydrogenation, the transversely oriented pentacene molecules connected with the vertically oriented ones, resulting in the formation of T-shaped pentacene junctions (Fig. 2c).

Further analysis of our STM images revealed that pentacene molecules traversing the Au grooves exhibited a dumbbell-like feature, where the central sites of the benzene rings appeared darker, while the terminal sites were brighter. This diversity in the observed T-shaped configurations may stem from two factors. Firstly, the breaking of the planar geometry is attributed to the interaction between the substrate surface and the molecules. Secondly, when pentacene molecules crossed the Au grooves, the central region of the pentacene molecule was positioned lower than the benzene ring, potentially enhancing the interaction at the central benzene ring. As a result, this interaction facilitated the formation of a carbon five-membered ring connection, along with the creation of chiral pentacene T-shaped junctions (Fig. 2d).

## Multi-configurations of the picene dimer

In addition to pentacene molecule, we explored the influence of different precursors by employing picene, an isomer of pentacene. Picene molecules consist of five benzene rings alternating in an armchair pattern, forming a structure resembling the Olympic rings. In contrast, pentacene molecule have their five benzene rings arranged linearly, representing distinct spatial arrangements of same atoms. Our aim was to investigate how the precursor type affects the dehydrogenation and polymerization processes on the Au(110) surface and their impact on the resulting chain structure. In our experiments, we deposited 0.8 ML of picene molecules onto an Au(110) substrate at room temperature (Fig. 3b). Interestingly, we observed that over 90% of the picene molecules underwent self-assembly, similar to the self-assembly behavior of pentacene (Fig. 3a). However, a notable distinction from pentacene that the majority of picene

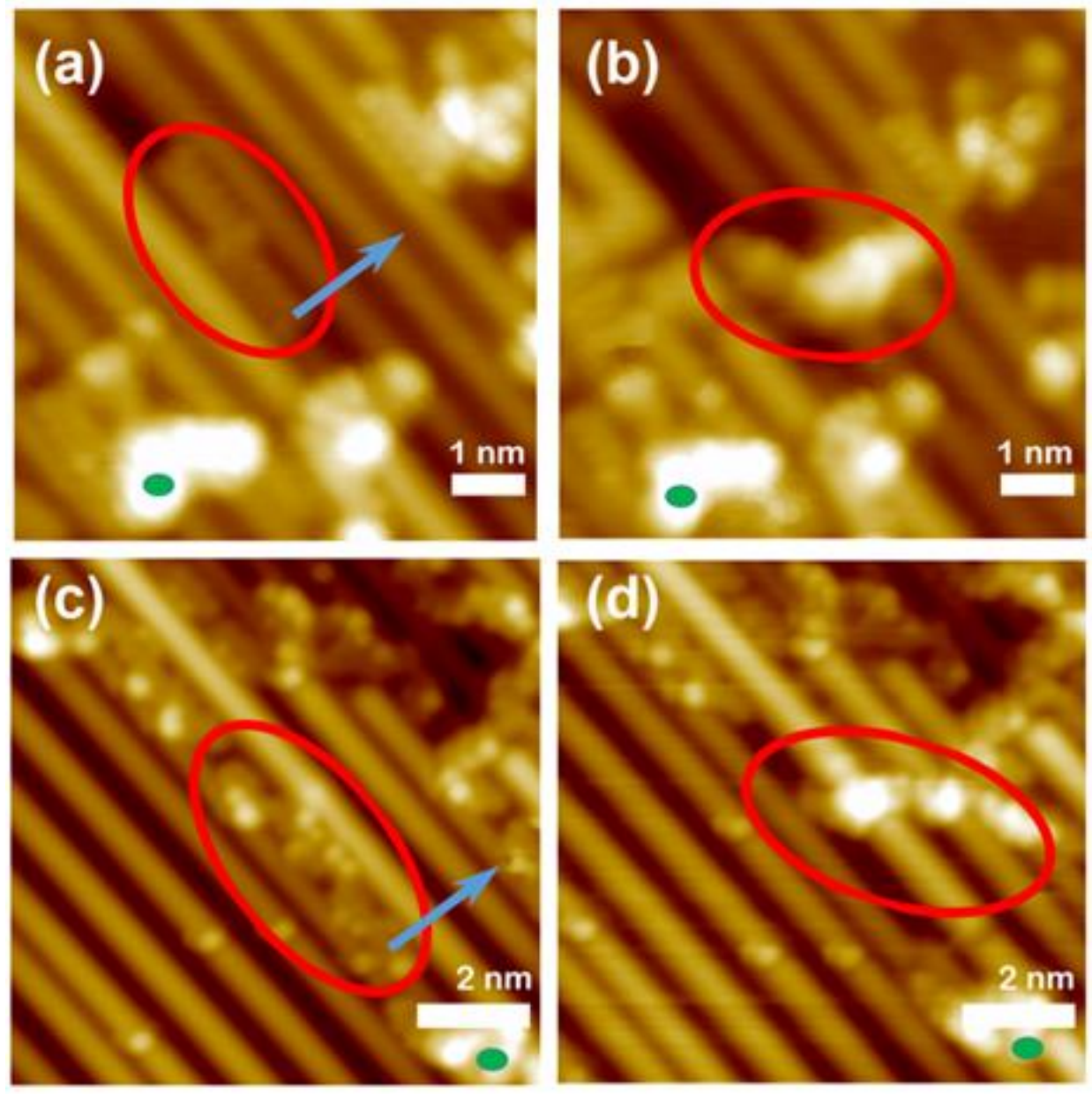

**Fig. 4** Pentacene dimer (a) before and (b) after STM tip manipulation ($V_s$ = −2 V and $I$ = 2 nA). Picene dimer (c) before and (d) after STM tip manipulation. ($V_s$= −2.5 V and $I$ = 1 nA)

molecules began to desorb from the surface at 576 K, which is lower than the desorption temperature observed for pentacene.

Notably, a distinct variations in the chemical reactions by Au(110) substrate 催化 were observed when comparing picene to pentacene. Firstly, the picene chains exhibited a significantly higher occurrence of superimposed configurations, despite these types of chains follow the connect junction through a carbon four-membered ring. Second, although the molecular chain length for the picene precursor is closely resembled chain from the pentacene precursor, the angles between chain segments were different (Fig. 3a). Specifically, the bond angles between chain segments formed by picene molecules exhibit variety (Fig. 4e), leading to diverse non-linearity one-dimensional graphene-like molecular chains.

The diversity of graphene-like molecular chains with different precursors is attributed to the molecule configuration. The picene with the benzene rings alternating in armchair pattern leads to a relatively higher energy state within the Au grooves and result in weaker interaction between the Au substrate and the molecule. Consequently, the templating effect of the Au grooves on picene molecule is minimal. Under specific annealing temperatures, this weak interaction prevents picene molecules from adopting a fixed orientation within the Au grooves, thereby hindering the formation of well-defined, straight-chain molecules after chemical reactions. In contrast, pentacene precursor molecules exhibit stronger interaction with the Au substrate, resulting in a significantly enhanced templating effect.

## Manipulation of the pentacene/picene dimers

Lastly, we utilized STM tip manipulation[22] to extract the dimer chains from the gold grooves. The tip approached near the sample at a distance of approximately 0.5 nm, with a bias voltage set at 3 V. The tip was then moved into the direction shown by the blue arrow (Fig. 4c). Remarkably, the entire chain preserved its original length without splitting into two separate sections. This strongly indicates the presence and picene of a robust four-membered ring connection between the two pentacene and picene segments, rather than the existence of two independent molecules without a connection or with weak single-bond connections. Moreover, compared with the dimer formed by the precursor of picene, it seems that the dimer formed by the precursor of picene is easier to push out from the Au(110) groove by STM tip. And the dimer formed by the precursor of picene is easier to maintain a straight chain configuration, while the dimer formed by pentacene is easy to fold after being pushed out of groove. This demonstrates the constructing targeted molecular chains with diverse configurations using tip modification. Our approach holds significant promise for high-precision applications in novel nanodevices, paving the way for innovative developments in nanoscale engineering.

## Conclusions

In summary, we successfully fabricated and synthesized multiple graphene-like nanoribbons using different precursors (pentacene and picene) constrained on a 1D Au(110) substrate. Our findings confirmed that the pentacene and picene monomers were connected via four- or five- membered ring. Furthermore, we investigated the system's mechanism by variating annealing time and substrate temperature. Additionally, we exhibited the technique to manipulate pentacene and picene dimers via a STM tip, showcasing a promising approach for the precise fabrication of novel-shaped nanodevices.

## Author contributions

H.Z. designed and supervised the work. Y.Y., Y.W. S.W. performed the experiments with assistance from Q.Z., H.B.Z, J.Z.; Y.Y., Y.W. and H.Z. analyzed the data and wrote the manuscript. All authors contributed to the scientific discussion and manuscript revisions.

## Conflicts of interest

There are no conflicts to declare.

## Data availability

No primary research results, software or code have been included and no new data were generated or analysed as part of this review.

## Acknowledgements

This research was supported by the National Natural Science Foundation of China (grant no. 12074357), the Innovation Program for Quantum Science and Technology (grant no. 2021ZD0302800), Anhui Provincial Key Research and Development Project (grant no. 2023z04020008), the CAS Project for Young Scientists in Basic Research (grant no. YSBR-046), the Fundamental Research Funds for the Central Universities (grant no. WK9990000118, WK2310000104). Thanks to Prof. Changgan Zeng, Prof. Ping Cui, Hongbin Zhu and Jie Zhou of the University of Science and Technology of China for valuable help and fruitful discussions.